\documentclass[twocolumn,times,tighten]{aastex631}

\usepackage{natbib}
\usepackage{graphicx}

\usepackage{ulem}

\newcommand{\del}[1]{\textcolor{magenta}{{\iffalse{#1}\fi}}}

\shorttitle{PSR J1951+2837 observed by the LPA and FAST}
\shortauthors{T.V. Smirnova et al.}

\begin{document}

\title{A nearby pulsar J1951+2837 observed by the LPA and FAST}
   
\author{T.V. Smirnova}
\affiliation{Lebedev Physical Institute, Astro Space Center, Pushchino Radio Astronomy Observatory, 
142290, Radioteleskopnaya 1a, Moscow reg., Pushchino, Russia}
   
\author[0000-0002-6423-6106]{D.~J. Zhou}
\affiliation{National Astronomical Observatories, Chinese Academy of Sciences, Jia-20 Datun Road, ChaoYang District, Beijing 100012, China}
\affiliation{State Key Laboratory of Radio Astronomy and Technology, Beijing, China, Beijing 100101, China }

\author{M.A. Kitaeva}
\affiliation{Lebedev Physical Institute, Astro Space Center, Pushchino Radio Astronomy Observatory, 
142290, Radioteleskopnaya 1a, Moscow reg., Pushchino, Russia}

\author[0009-0002-5322-3964]{S.A. Andrianov}
\affiliation{Lebedev Physical Institute, Astro Space Center, Pushchino Radio Astronomy Observatory, 
142290, Radioteleskopnaya 1a, Moscow reg., Pushchino, Russia}

\author[0000-0002-9274-3092]{C. Wang}
\affiliation{National Astronomical Observatories, Chinese Academy of Sciences, Jia-20 Datun Road, ChaoYang District, Beijing 100012, China}
\affiliation{School of Astronomy and Space Science, University of Chinese Academy of Sciences, Beijing 100049, China}
\affiliation{State Key Laboratory of Radio Astronomy and Technology, Beijing, China, Beijing 100101, China }

\author[0000-0002-6437-0487]{P.~F. Wang}
\affiliation{National Astronomical Observatories, Chinese Academy of Sciences, Jia-20 Datun Road, ChaoYang District, Beijing 100012, China}
\affiliation{School of Astronomy and Space Science, University of Chinese Academy of Sciences, Beijing 100049, China}
\affiliation{State Key Laboratory of Radio Astronomy and Technology, Beijing, China, Beijing 100101, China }

\author[0000-0002-9274-3092]{J.~L. Han}\thanks{E-mail: hjl@nao.cas.cn}
\affiliation{National Astronomical Observatories, Chinese Academy of Sciences, Jia-20 Datun Road, ChaoYang District, Beijing 100012, China}
\affiliation{School of Astronomy and Space Science, University of Chinese Academy of Sciences, Beijing 100049, China}
\affiliation{State Key Laboratory of Radio Astronomy and Technology, Beijing, China, Beijing 100101, China }

\author[0000-0003-0042-0884]{S.A. Tyul'bashev}\thanks{E-mail: serg@prao.ru}
\affiliation{Lebedev Physical Institute, Astro Space Center, Pushchino Radio Astronomy Observatory, 
142290, Radioteleskopnaya 1a, Moscow reg., Pushchino, Russia}

\date{Received ; accepted}

\begin{abstract}
PSR J1951+2837 is a nearby pulsar with a period of 7.334~s and dispersion measure of DM = $2.9\pm0.6$~pc~cm$^{-3}$, located about 200 or 300~pc from the Sun. It occasionally radiates bright pulses and has been observed by the Large Phased Array (LPA) radio telescope at 110~MHz and by the Five-hundred-meter Aperture Spherical radio Telescope (FAST) at 1250 MHz. We detected only 343 pulses in 228 LPA observation sessions and 5 bright pulses in two FAST sessions. 
Based on the times of arrival (TOAs) of these bright pulses, we determined the coherent timing solution for this pulsar at a frequency of 110 MHz. 
Based on flux densities ($S$) of these bright pulses at two frequencies ($\nu$), we found that it is probably one of the known pulsars with the lowest luminosities to date, with a spectral index of about $\alpha = (2.5 - 3.2)$ for $S\sim \nu^{-\alpha}$. 
\end{abstract}


\section{Introduction}

Since the discovery of pulsars in 1967 \citep{Hewish1968}, dozens of surveys have been conducted to find more pulsars. There have already been more than 3,700 pulsars in the Australia Telescope National Facility (ATNF) pulsar catalog\footnote{https://www.atnf.csiro.au/research/pulsar/psrcat/}  \citep{Manchester2005}. Despite the large number of surveys already conducted, new surveys are still going on by, for example, \citet{Sanidas2019, McEwen2020, Krishnan2020, Han2021, Amiri2021, Tyulbashev2022, Bhat2023, Han2025}, mainly because the newly conducting surveys have the improved sensitivities due to the enlarged observation frequency band and due to the commissioning of new radio telescopes with large effective areas, and new process methods. These studies are usually motivated to find new types of pulsars for a better understanding of their populations in the Milky Way. 

New surveys are often conducted in the sky areas that have been repeatedly searched in early surveys. Consequently, these new surveys detect pulsars that were missed in previous searches. There may be pulsars whose flux densities are below the detection threshold for early surveys, or new pulsars with some features that caused them to be missed previously. For example, they can be nulling pulsars or rotating radio transients \citep[RRATs,][]{rrat06}, with sporadic pulses with strong intensity variability, or pulsars with very long periods \citep{Zhou2023}  which could not be detected in the periodical searches, and other unknown reasons. The radiation characteristics of these unusual pulsars are important for understanding the physics of processes in the pulsar's magnetosphere.

The Large Phased Array (LPA) radio telescope has been used to search for long-period transients with periods from 2 to 90~s by using the Fast Folding Algorithm (FFA) \citep{Tyulbashev2024}, and can achieve a sensitivity of 1~mJy at a frequency of $\nu =110$ MHz. PSR J1951+2837 (i.e. J1951+28) was discovered in the sky area of the declinations of $+21^o< \delta < +42^o$ \citep{Tyulbashev2024}.

\begin{figure*}
   \centering
   \includegraphics[width=0.75\textwidth]{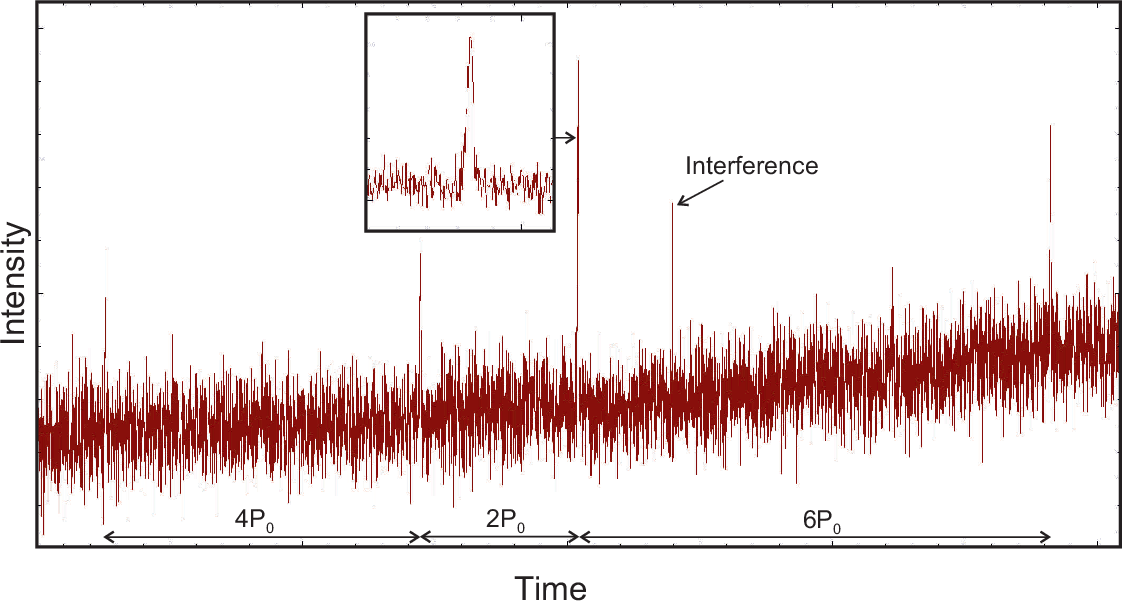}
   \caption{Raw data for four pulses of PSR J1951+2837 recorded directly by the LPA on January 7th, 2019, summed without dispersion compensation, gain equalization, and interference mitigation. The inset shows the details of the strongest pulse.}
    \label{fig:fig1}%
\end{figure*}

In this paper, we present the detailed results of the long-period pulsar, PSR J1951+2837, by the LPA and the Five-hundred-meter Aperture Spherical Telescope \citep[FAST,][]{fast06}. This pulsar has a period of $P_0=7.334$~s. In the ATNF catalog for October 2024, there are only 6 radio pulsars and 3 RRATs having a period $P_0 > 7.3$~s. In Section 2, observations and their processing are briefly introduced. In Section 3,  the results and discussion are presented. Conclusions are given in the last section.

\section{The LPA and FAST observations, data processing}

\subsection{The LPA Observations}

PSR J1951+2837 was first detected by the LPA \citep{Tyulbashev2024} at the right ascension  $\alpha_{2000}=19^h51^m30^s\pm 45^s$ and the declination $\delta_{2000}=28^o46{'}\pm10{'}$. It is located in the Galactic plane at ($l,b$) = ($64^{\circ}.952, b = 0^{\circ}.889$), with a dispersion measure $DM=3.5\pm 2.0$ pc cm$^{-3}$.

Naturally, PSR J1951+2837 falls into one of the LPA beams every day, and the LPA records data of PSR J1951+2837 daily at a frequency of 110.4 MHz. The LPA has an observational bandwidth of 2.5~MHz, and it is split into 32 frequency channels (each with a channel width of 78~kHz). Data of each channel are sampled with a time resolution of 12.5~ms \citep{Tyulbashev2022}. An example of recording raw data J1951+2837 is shown in Figure~\ref{fig:fig1}, with the duration of 88.01~s between the two outmost pulses.

The processing of the data obtained by the LPA consists of two stages. In the years from 2014 to 2024, we selected these days with an average profile of signal-to-noise ratio $(S/N) \geqslant 6$. There are data for 12 such days out of about 3,000 days. Visually examining data of these days, we find that 3 to 5 individual pulses are recorded in the daily session lasting 3.5 minutes. In total 46 such pulses were detected. On average, the detection rate of pulses in these 12 sessions is about 1 pulse per minute.

For further pulse analysis, data slices of 32 frequency channels around the detected pulses with a length of 101 recorded data samples are selected. First, the baseline was subtracted carefully for each frequency channel. Second, the standard deviations are determined for each channel by using data outside the pulse range, and then they are used to normalize the data amplitude for all channels. A search is finally made to dedisperse data from all channels and sum them for the maximum amplitude of the pulses, and the best profile is produced. 

The dispersion measure of PSR J1951+2837 has been estimated in \citet{Tyulbashev2024} which is DM=3.5$\pm$2.0 pc\,cm$^{-3}$. The low accuracy of DM determination is mainly caused by the wide average profile with a width of  $W_{0.5}$ = 130~ms (10 sampling points) and a small dispersion shift over a 2.5 MHz frequency band: only three sampling data points for the change of $DM$ in 2.5~pc~cm$^{-3}$.

\begin{figure*}
   \centering
   \includegraphics[width=0.85\columnwidth]{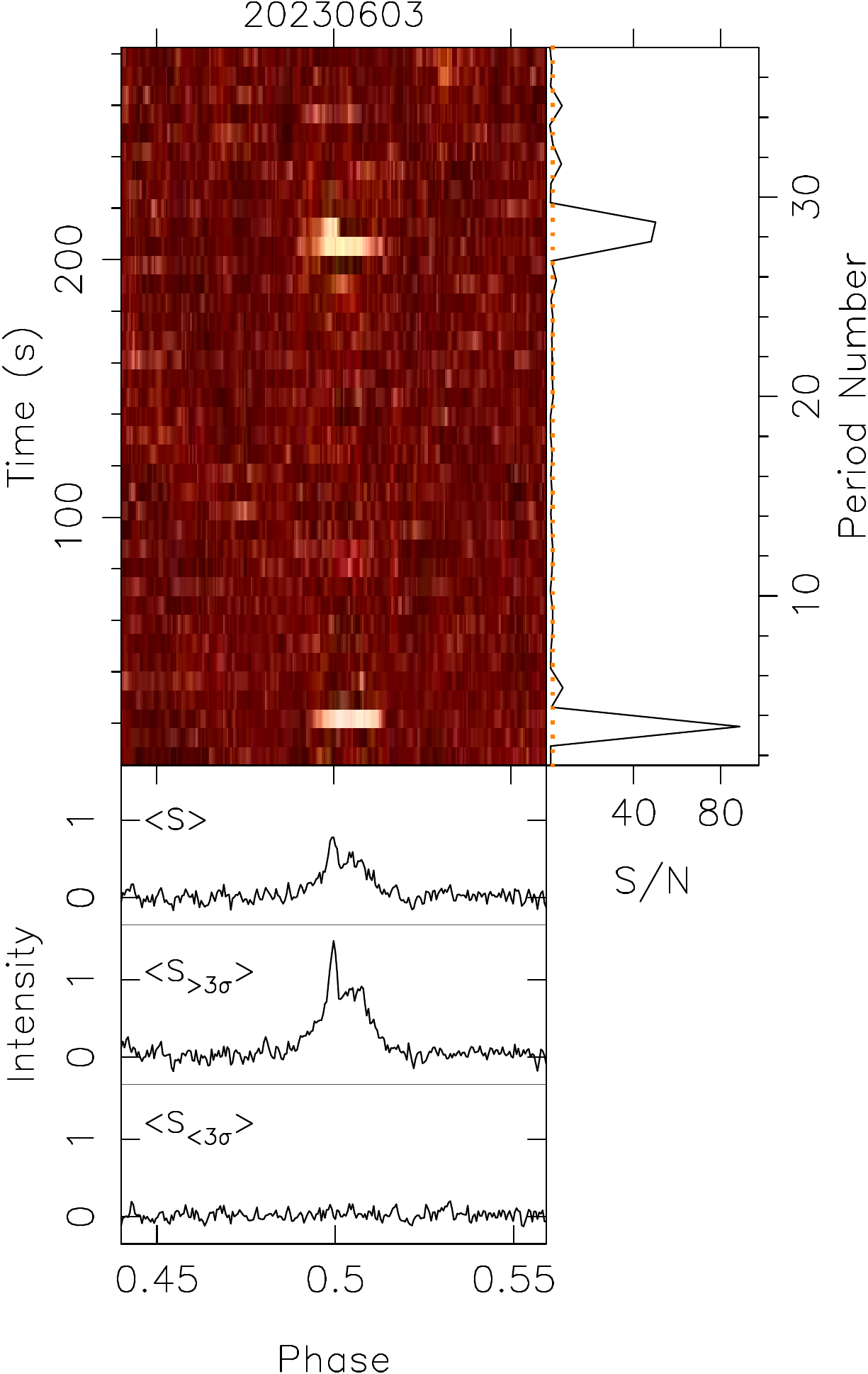} 
   \hspace{5mm}
   \includegraphics[width=0.85\columnwidth]{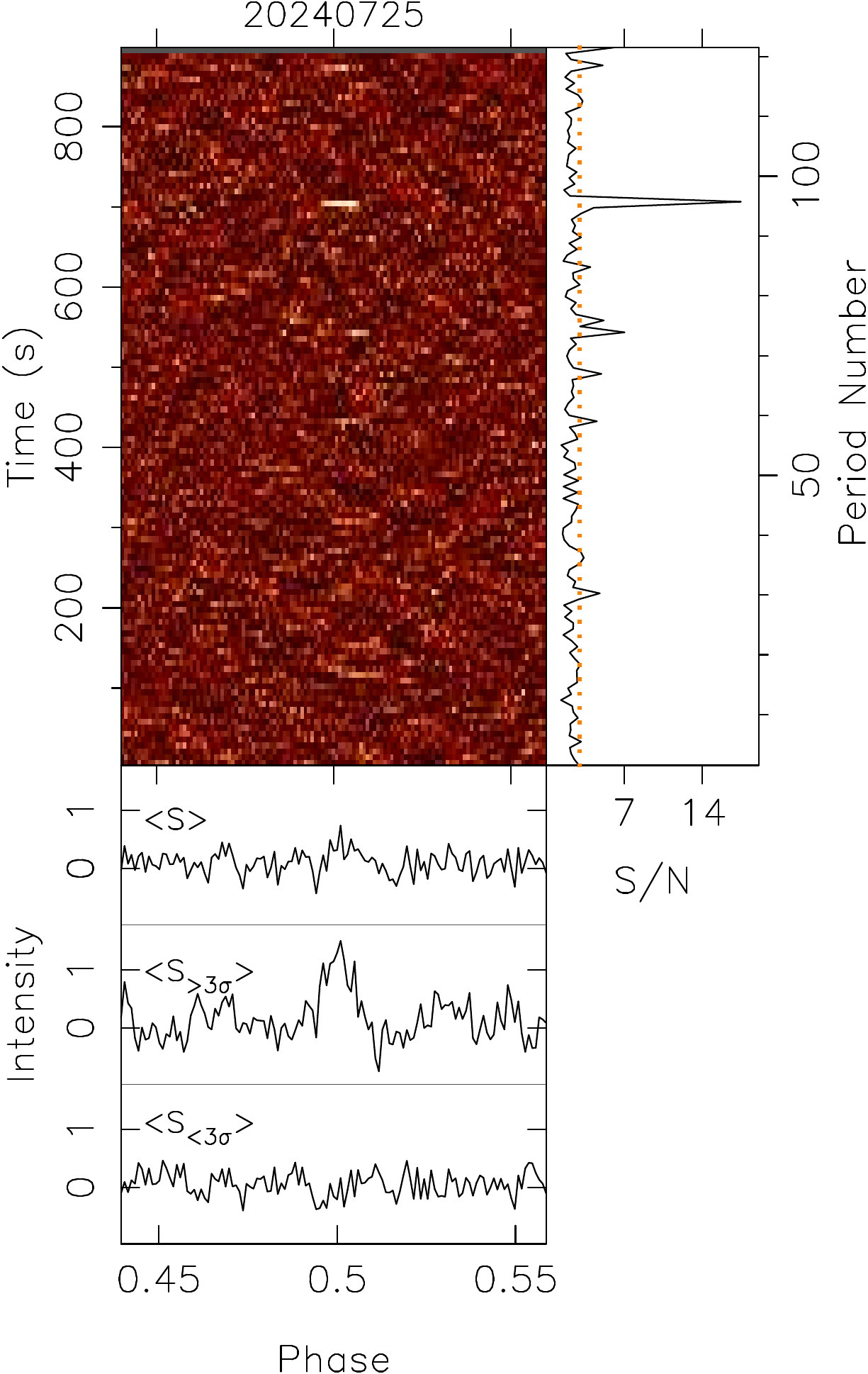}
   \caption{The pulse stacks obtained from the FAST verification observations on 20230603 and 20240725. The brighter color reflects a higher intensity. The right subpanel shows the observed pulse amplitude in S/N units, with a dotted line for 3$\sigma_n$ (root mean square of the noise). The profiles summed over the entire observations from pulses with $S/N > 3 \sigma_n$ or $<3 \sigma_n$ are shown in the bottom subpanels, with the peak centered on the rotation phase 0.5. 
   }
    \label{fig:fast}%
    \end{figure*}
\subsection{The FAST Observations}

The sky region around the initial position of PSR J1951+2837 has been observed twice by the FAST with the 19-beam L-band receiver 
by using first the snapshot observation mode \citep{Han2021} on June 03, 2023, and then by the tracking mode for 15 minutes on July 25, 2024, after the pulses from this pulsar have been detected. The FAST snapshot observation can have 76 beams to check the pulsar location, as shown in Figure 4 in \citet[]{Han2021} .
Both FAST observations are made at 1250~MHz with a bandwidth of 500~MHz, covered by 2048 frequency channels. For each channel, the data of four polarization channels of  $XX$, $YY$, Re[$X^{*}Y$], and Im[$X^{*}Y$] have been recorded in FITS files with a sampling time of 49.152 $\mu$s. 

The processing of FAST observations also consists of two stages. First is to find the pulses of PSR J1951+2837 in the recorded data of 76 beams by the FAST snapshot mode, 5 minutes for each beam. 
The pulsar did not appear in the periodical search but in the single pulse search. The coordinates of the beam center for the single pulse detection are $\alpha_{2000}=19^h51^m37.7^s$, $\delta_{2000}=28^o36'48''$, which are taken as the pulsar coordinates with an uncertainty of $\pm1.5'$. The second
FAST tracking observation was carried out by pointing the central beam of
the 19-beam receiver to this position. PSR J1951+2837 exhibits as an RRAT in FAST data. Bright pulses were detected during each of two FAST observation sessions, together with a different number of weaker pulses in the durations of 5 minutes and 15 minutes, respectively (see Fig.~\ref{fig:fast}).

\section{Results and discussion}

Careful examinations of the LPA and FAST data reveal properties of PSR J1951+2837.

\subsection{Estimation of the dispersion measure}

The DM value was estimated more accurately from the measured signal delay of narrow pulses on the spaced frequency channels: 
$\triangle t\simeq 4.15\times 10^6 \times (\nu_1^{-2} - \nu_2^{-2}) \times DM$ (ms),
where $\triangle t$ is the observed pulse shift in ms, $\nu_{1,2}$ are the channel frequencies in MHz. We have obtained more accurate DM estimates from the LPA and FAST observations of narrow bright pulses. 
From the observations on the LPA, two narrow pulses ($W_{0.5} \sim 25$ ms in the frequency channel) were used at spaced frequencies for May 14, 2021. Based on the shift of the profile peak in the frequency channels, we obtained the value DM=3.2$\pm{0.8}$ pc cm$^{-3}$.

From FAST observations, the best DM estimate is DM=2.5$\pm{0.9}$ pc cm$^{-3}$ obtained from the strongest pulse observed on June 03, 2023 which has a $S/N > 80$. Combining these two estimates, we obtained the DM being 2.9$\pm$0.6~pc~cm$^{-3}$. Using the DM value, we estimated the distance to the pulsar as $D\approx 200$~pc based on the YMW16 model for the electron density distribution in our Galaxy \citep{Yao2017} or 300~pc based on the NE2001 model \citep{ne2001}.

\begin{figure}
 \centering
 \includegraphics[width=0.9\columnwidth]{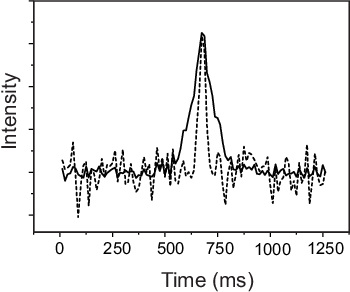}
 \caption{The integrated pulse profile (in arbitrary scale) obtained by adding 33 pulses observed by LPA presented by solid line, with $W_{0.5}$ = 100 ms. The dotted line shows one individual pulse detected on  May 14, 2021, having a width of $W_{0.5}$ = 37 ms.}
    \label{fig:fig2}%
    \end{figure}

\subsection{Pulse widths}

Observed individual pulses vary a lot at both frequencies. 
Individual pulses observed by LPA have widths from 37 ms to 100~ms, 
with typical 
widths of $W_{0.5}$= 60-80 ms. The mean pulse of 33 pulses with a signal-to-noise ratio (S/N) $> 5$ has a width of $W_{0.5}$ = 100 ms and a S/N of 33.6, as shown in Figure~\ref{fig:fig2} together with the narrowest (width 37 ms, dotted line, S/N=8.9). 

Individual pulses detected by FAST have widths from 20 to 224 ms. The width of the mean profile accumulated over 5 minutes on June 03, 2023 is  $W_{0.5}$ = 110 ms, and that from 121 periods over 15 minutes observed on July 25, 2024 is $W_{0.5}$ = 72 ms. 
%
Figure~\ref{fig:fast} shows pulses from the two FAST observation sessions. Along rotations 38 periods, we detect 14 pulses with $S/N>3$, 
4 pulses with $S/N>6$, 
3 pulses with 
$S/N>40$. It is these 3 brightest pulses that act as the signatures of  
this pulsar in the verification FAST observation session on June 03, 2023. 

We therefore conclude that the pulse width of the mean profiles at different frequencies should be between 70-130 ms. The duty cycle of this pulsar is $W_{0.5}/P_0$=0.016, very typical for slow pulsars. 

Since not many periods have been observed for this pulsar, the resulting total intensity profiles are dominated by several bright pulses. 
Therefore, it is hard 
to discuss the frequency dependence of the profile width according to the available data. 

\subsection{RRAT features}

FAST observations show that the emission features of PSR J1951+2837 are consistent with the definition for RRATs \citep{Zhou2023}. During the 280-second observation session, only 4 strong ($S/N>6$) pulses were detected from 38 periods. From 900-second data, only 8 strong pulses were detected from 121 periods. 
When excluding four strong pulses recorded on June 03, 2023, the accumulated profile of 24 pulses with an amplitude less than $3\sigma_n$ is not visible at the peak phase of 0.5 (Figure~\ref{fig:fast} bottom panel). That is the feature of a regular RRAT \citep{Zhou2023}. On the other hand, the summation of FFA spectra of the LPA observations over 5 years (3.5 min per day or $\approx 7.2\times 10^5$~s) has a periodic signal {\bf with $P=7.3342$ of the strongest FFA harmonic} \citep{Tyulbashev2024}. This means that very weak pulses can still be detected at the radio frequency of 110 MHz. 

\begin{figure}
   \centering
 \includegraphics[width=0.99\columnwidth]{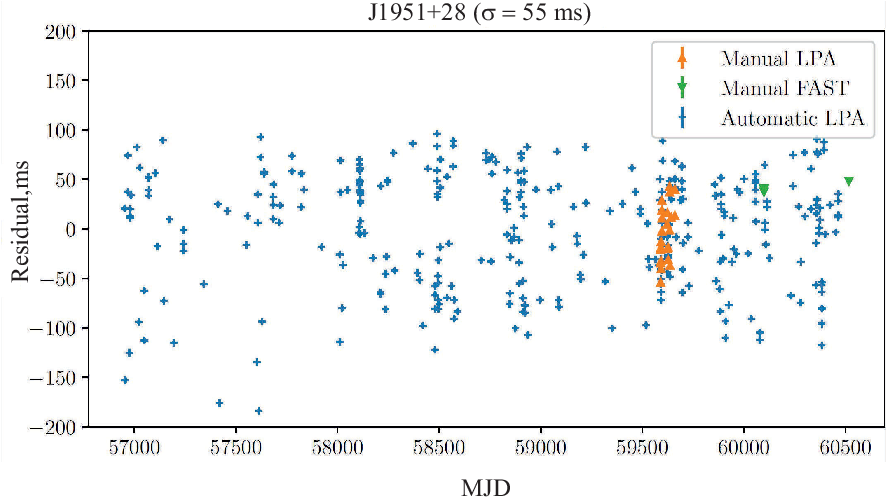}  
   \caption{The residual of TOAs of pulses detected by LPA and FAST after fitting the timing solution. The vertical axis shows the residual deviations in ms, and the horizontal axis is the time in modified Julian Days (MJD).
   }
\label{fig:fig3a}%
\end{figure}

\subsection{Timing by LPA}

After PSR J1951+2837 was detected at FAST and its coordinates were determined with an accuracy of 1.5$^\prime$, we did pulsar timing based on the method developed recently by \citet{Andrianov2025}. 

Early attempts to do pulsar timing in the frame of our searching program were unsuccessful, and we attributed this failure to the fact that a quartz oscillator has poor accuracy for controlling our sampling. However, a recent study \citet{Andrianov2025} showed that timing can be done with a correction of the time arrival of pulses by using observations of strong pulsars with known $P$ and $\dot P$ values and located in approximately the same declinations as our source. The data of these pulsars were used as intermediate reference clocks, which made it possible to qualitatively improve the accuracy of determining the time of arrival of pulses (TOA). As shown in \citet{Andrianov2025}, the typical accuracy of determining the period can go to the 9th decimal digit, which has been verified by 24 pulsars.

PSR J1951+2837 has a low $DM$, a wide mean profile, and relatively small S/N, so it is difficult to distinguish pulses from interference. It was assumed that the pulses radiated from a pulsar have time intervals related to the pulsar period (see Fig.1). By using individual pulses observed on different days, the 
peaks of individual pulses can be used for timing. 
For PSR J1951+2837, the period has been determined up to a few units in the fourth decimal digit. This accuracy in determining the period leads to the phase loss for the intervals of $>$3 days. To restore a more accurate value of the period, time intervals were chosen in which pulses were observed for 2 consecutive days as a reference. By analyzing the previously confirmed pulses, the pulsar period has been refined so that the pulse phase can be maintained for about 1 month. Based on the updated value of the period and the coordinates of the pulsar, which are known with an accuracy of $\pm 1.5^\prime$, we conducted a repeated search for weak pulses over 10 years. By iteratively searching for pulses and specifying the parameters of this pulsar, we have achieved the phase-connected solution for all pulses throughout the entire observation time over 10 years. The standard TEMPO2 timing program \citep{Hobbs2006} was then used for the parameter adjustment. $DM$ was fixed and not specified during the timing. We used a few known pulsars as intermediate reference clocks to improve TOA for PSR J1951+2837.

Figure~\ref{fig:fig3a} shows the residual deviations from the time of arrival (TOAs) of pulses after timing-fitting. 
By selecting $P$, $\dot P$ and the coordinates of the pulsar together in TEMPO2, we obtained residuals similar to white noise. In Figure~\ref{fig:fig3a}, the pulses for February 2022 are highlighted in orange for two or more pulses detected per observation session. The green triangles show data points from FAST observations. After compensating for the constant phase difference, the residuals of pulses fell within the limits of the normal residuals of the LPA telescope.

\begin{table}
\centering
\caption{Basic parameters of PSR J1951+2837. The digits in the brackets are 1$\sigma$ uncertainties for the last digit. The TCB scale and the ephemeris DE405 were used for timing.
}
\label{tab:parameters}%
\renewcommand\arraystretch{0.8}
\begin{tabular}{lr}
\hline
 Parameters  & Values \\
 \hline
 Right Ascension (J2000)      &  $19^h51^m37.65(40)^s$   \\
 Declination (J2000)             & $+28^{\circ}37^{\prime}43(6)^{\prime \prime}$   \\
 Galactic Longitude ($^{\circ}$) &  64.952     \\
 Galactic Latitude  ($^{\circ}$) &  0.889    \\
 Reference epoch for period (MJD)            &  60713   \\
 Spin period, $P$ (s)                 &  7.3339006895(11)  \\
 Spin period derivative, $\dot{P}$ ($10^{-14}$~s\,s$^{-1}$) & 2.93179(61)  \\
 Characteristic age, $\tau_{\rm c}$ (Myr) & 4 \\
 Surface magnetic field, $B_{\rm s}$ (10$^{13}$ G) & 1.5 \\
 Spin-down luminosity, $\dot E$ ($10^{30}$) \, erg s$^{-1}$  & -2.9 \\
 DM ($\rm pc\,cm^{-3}$)  &  2.9$\pm$0.6     \\
 RM ($\rm rad\,m^{-2}$) & $-$2.2$\pm$0.7      \\
 Spectral index $\alpha$ for $S\sim \nu ^{-\alpha}$    &  2.5 to 3.2      \\
\hline
\end{tabular}
\end{table}

From a total of 3542 LPA observation sessions, corresponding to approximately 200 hours of observation, we detected in total 343 pulses from 228 sessions. The median and average values of the pulses half-width are 135 and 133 ms with $\sigma=22.2$~ms. The strongest pulse was detected on August 02, 2016, with a S/N=22.6. From the timing solution, the basic parameters (with uncertainty in brackets) of PSR J1951+2837 are presented in Table~\ref{tab:parameters}.
The timing solution enables us to estimate the surface magnetic field ($B$) and the characteristic age ($\tau$). Assuming a magneto-dipole radiation, we get $B=1.5 \times 10^{13}$~G, $\tau = 4 \times 10^6$~yr. 

According to the ATNF catalog, estimates of the age and magnitude of the magnetic field (for $P>7.3$~s) are known for 4 and 21 pulsars. Our estimates of $\tau_c$ and $B$ falls on the median values. Generally, this pulsar is close to the death-line in the $P-\dot P$ diagram and falls in to XDINs group of pulsars.

\begin{figure}
   \centering
   \includegraphics[width=0.8\columnwidth]{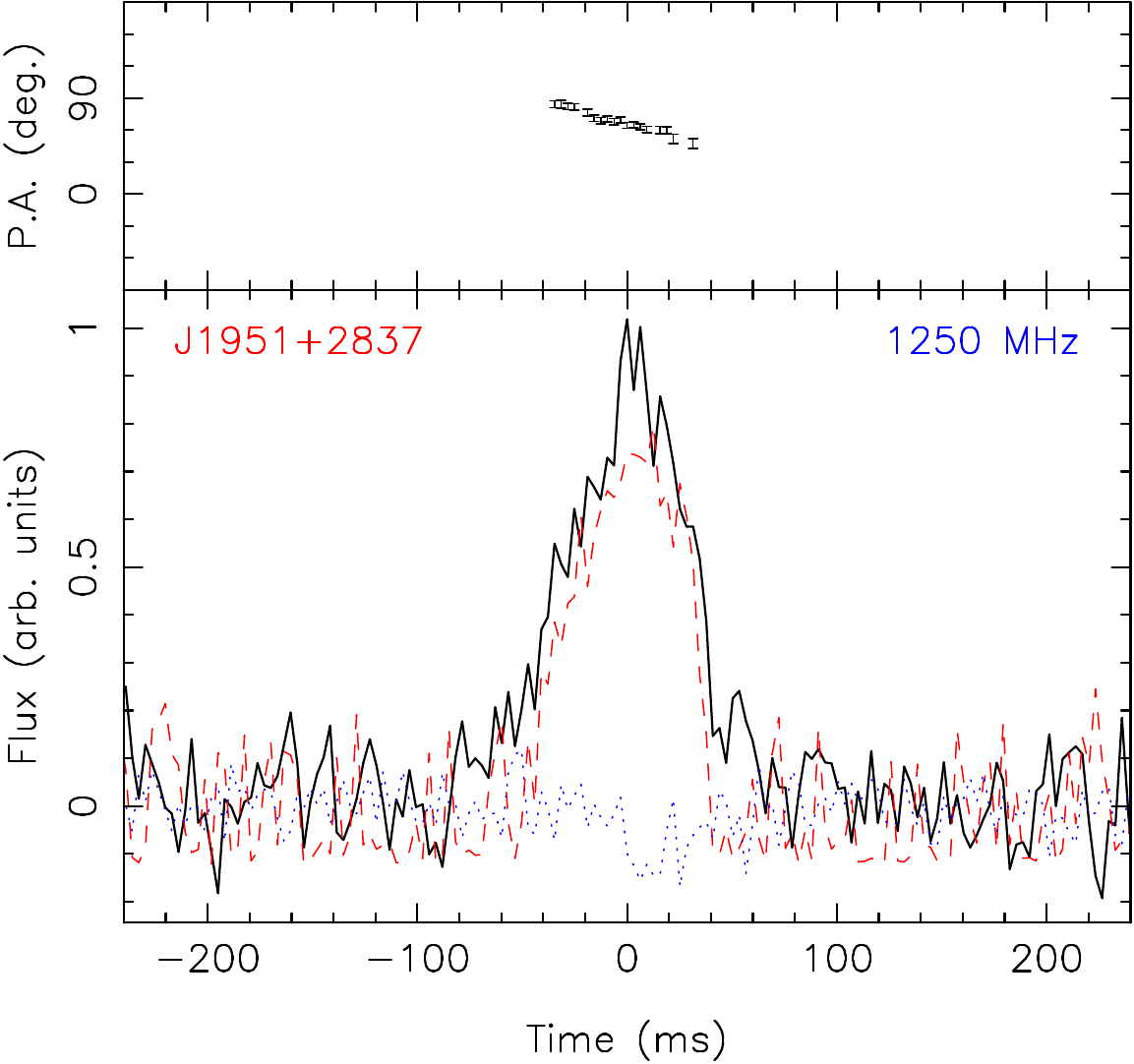}
   \caption{The polarization profiles for the brightest pulse detected by FAST on July 25, 2024. The total intensity, the linear polarization, and circular polarization profiles are given by the solid line, dashed line and dotted line in the bottom sub-panel. The polarization angles are shown in the upper sub-panel. 
   }
    \label{fig:fig5}%
    \end{figure}
\subsection{Polarization}


In LPA observations, data of one linear polarization channel were recorded, and therefore no polarization characteristics of this pulsar can be determined at 110~MHz. 

FAST observation in the session of July \textbf{25}, 2024 has the polarization data recorded. Following the procedures in \citet{Zhou2023} and \citet{wang2023},  we got the polarization profiles of the brightest pulse, as shown Figure~\ref{fig:fig5}. They show a high linear polarization degree of 85.3$\pm$4.2\%, and a circular polarization degree of 5.0$\pm$1.3\%. From the polarization data, we derive the rotation measure (RM) of PSR J1951+2837 as being $RM = -2.2\pm 0.7$ rad~m$^{-2}$.

Using the observed DM and RM of PSR J1951+2837, one can estimate the parallel component of the magnetic field in the interstellar medium between the pulsar and the Sun \citep{han06}:
$ <B_{||}> = 1.23 \, \mu \, G \left( \frac{RM} {\rm rad \, m^{-2}} \right) \left(\frac{\rm DM}{\rm pc \, cm^{-3}}\right)^{-1} = -0.9 \times 10^{-6}$~G,
which 
is a normal value in the interstellar magnetic field in the Galaxy \citep{han18,xu22}, with the direction toward the pulsar. 

\subsection{Flux density,  spectral index and luminosity}

We have observed pulses of PSR J1951+2837 at two bands, and the spectral index of the pulsed emission, $\alpha$, can be estimated from the flux densities via $S\sim \nu^{-\alpha}$. The maximum integrated flux density at  110 MHz can be estimated from several strong pulses detected by LPA on May 14, 2021, that is $S_{\rm int}$ = 11 mJy (as shown in \cite{Tyulbashev2024}). 
The mean flux densities at 1250 MHz from the two FAST observation sessions are $S_{\rm int}$=0.0273 mJy (June 3, 2023) and $S_{\rm int}$=0.0052 mJy (July 25, 2024), which are estimated with the system noise ($\sigma$) by the function:
\begin{equation}
    \sigma = \frac{T_{\rm sys}}{G_0 \sqrt{n_{\rm p}T_{\rm obs}BW}},
\end{equation}
and the flux density is:
\begin{equation}
    S_{\rm int} = \frac{\sum S_{\rm i}}{\sigma_{\rm off}\sqrt{n}}\sigma\sqrt{\frac{W}{P-W}},
\end{equation}
hear the $T_{\rm sys}\sim$20K is the system noise temperature, $G_0$=16.1 K/Jy is the effective gain of the telescope \citep{JiangP2020RAA....20...64J}, $n_{\rm p}$=2 is the number of polarization summed, $T_{\rm obs}$ is the observation time duration, BW= 437.5 MHz is the frequency bandwidth after removed the upper and lower sidebands of 31.25 MHz \citep{Zhou2023}, the $\sigma_{\rm off}$ is the standard deviation of the off-pulse region of the average pulse profile, $\sum S_{\rm i}$ the sum of the points in the on-pulse region where data point value $S_{\rm i} > 3\sigma_{\rm off}$, and $n$ is the corresponding number of points,  $W$ is the on-pulse width ($n$ points), and the $P$ is the pulse width (total number of folded points). One can roughly combine these flux densities and get an estimate of the maximum $\alpha$ = 2.5 - 3.2, or $<\alpha>$=2.85, in the frequency range of 110 -- 1250~MHz. 

FAST observations were carried out in a frequency band from 1000 to 1500~MHz. It is also possible to estimate the in-band spectral index based on observations of strong pulses. We divided the 500 MHz band into 4 subbands, 125 MHz each, and got the flux densities in these subbands. The fitted spectral index for the averaged flux densities of bright pulses is  $<\alpha>$=2.5, similar to the values obtained from the mean pulse profiles in the two FAST sessions.

With the estimated spectral index, say $\alpha$=2.9, one can estimate the pseudo-luminosities of pulsars, $L=S \times D^2$, by assuming the beam-factor of 1/4$\pi$ \citep{Lorimer2004}). 
The distance of this pulsar is 
208~pc from the YMW16 model \citep{Yao2017} or 300~pc by the NE2001 model \citep{ne2001}. Hence the luminosity at a frequency of 110 MHz $L_{110}= S_{int} \times D^2$ = 0.48 mJy kpc$^2$ (for D =208~pc) or 0.99~mJy~kpc$^2$ (for D = 300~pc). Conventionally, pulsar luminosities are estimated for values at 400 MHz, for example, the luminosities of 748 pulsars in the ATNF pulsar catalog (as of October 2024). We used the spectral index to convert, and got $S_{400} =(110/400)^{2.9} \times 11 = 0.28$~mJy, and then $L_{400} = 0.28 \times 0.208^2$ = 0.012 mJy kpc$^2$, or $0.28 \times 0.300^2$= 0.025 mJy~kpc$^2$. This luminosity value is 37 times lower than the minimum luminosity at 400 MHz found in the ATNF catalog for the pulsar PSR J0307+7443 that is $L_{400}$=0.44 mJy kpc$^2$. If we take the spectral index of $\alpha$=2.5 obtained by FAST, we get $S_{400}$=0.44 mJy and $L_{400}$=0.019 mJy kpc$^2$, a luminosity 23 times lower than the minimum luminosity of PSR J0307+7443 at 400 MHz.

One may estimate the luminosity at a frequency of 1250 MHz based on the integral flux density obtained on FAST: $L_{1250} = 0.0273$ or $0.0052 \times 0.208^2 = 0.0011 - 0.0002$ mJy kpc$^2$. In the ATNF pulsar catalog, there are luminosities at 1400 MHz for 2442 pulsars, and the minimum luminosities are $L_{1400}$=0.0030 mJy kpc$^2$ for PSR J1105$-$4354 and or $L_{1400}$ =0.0024 mJy kpc$^2$ for PSR J1107$-$5907. Again, one can see that the luminosity of PSR J1951+2837 is several times lower than those of previously known pulsars. 

\section{Conclusion}

We have observed PSR J1951+2837 by using the LPA and FAST, and revealed properties of this pulsar. 

PSR J1951+2837 behaves as an RRAT, with one or two strong pulses every two minutes seen in 
FAST observations. 
In LPA observations, the pulsar was found to be a source of very weak periodic radiation \citep{Tyulbashev2024}. 

The timing solution was found from the long-term data recording from the LPA. We derived the 
period and the period derivative of PSR J1951+2837 and determined the its position. We determined the best $DM = 2.9\pm0.6$~pc~cm$^{-3}$ from the LPA and FAST observations, which indicates its proximity of only about 200~pc or 300~pc according to the electron density models. The luminosity of this pulsar is found to be the lowest among any known pulsars. From the FAST observation, we found that the pulses are a highly polarized, and determined the Faraday rotation of this pulsar as being $RM = -2.2\pm 0.7$ rad~m$^{-2}$. 
We also estimated the spectral index about $\alpha = 2.9$ in the range of 110-1250 MHz. 



\section*{Acknowledgment}
The Pushchino pulsar team is grateful to the LPA technical team for the monitoring data used in this paper. The authors thank L.B. Potapova for her help in the preparation of the paper. TVS, MAK and SAT were supported by the Russian Science Foundation no. 22-12-00236-$\Pi$ (https://
rscf.ru/ project/ 22-12-00236-$\Pi$/ ). 
This work utilized data from FAST (https://cstr.cn/31116.02.FAST). FAST is a Chinese national mega-science facility, built and operated by the National Astronomical Observatories, Chinese Academy of Sciences. 
The Chinese authors are supported by the Natural Science Foundation of China: No. 12588202, 11833009, and the National SKA Program of China 2020SKA0120100. 
%

\bibliographystyle{raa}
\bibliography{serg1} 

@ARTICLE{Hewish1968,
       author = {{Hewish}, A. and {Bell}, S.~J. and {Pilkington}, J.~D.~H. and {Scott}, P.~F. and {Collins}, R.~A.},
        title = "{Observation of a Rapidly Pulsating Radio Source}",
      journal = {\nat},
         year = 1968,
        month = feb,
       volume = {217},
       number = {5130},
        pages = {709-713},
          doi = {10.1038/217709a0},
       adsurl = {https://ui.adsabs.harvard.edu/abs/1968Natur.217..709H}
}

@ARTICLE{Manchester2005,
       author = {{Manchester}, R.~N. and {Hobbs}, G.~B. and {Teoh}, A. and {Hobbs}, M.},
        title = "{The Australia Telescope National Facility Pulsar Catalogue}",
      journal = {\aj},
         year = 2005,
        month = apr,
       volume = {129},
       number = {4},
        pages = {1993-2006},
          doi = {10.1086/428488},
archivePrefix = {arXiv},
       eprint = {astro-ph/0412641},
 primaryClass = {astro-ph},
       adsurl = {https://ui.adsabs.harvard.edu/abs/2005AJ....129.1993M}
}

@ARTICLE{Sanidas2019,
       author = {{Sanidas}, S. and {Cooper}, S. and {Bassa}, C.~G. and {Hessels}, J.~W.~T. and {Kondratiev}, V.~I. and {Michilli}, D. and {Stappers}, B.~W. and {Tan}, C.~M. and {van Leeuwen}, J. and {Cerrigone}, L. and {Fallows}, R.~A. and {Iacobelli}, M. and {Orr{\'u}}, E. and {Pizzo}, R.~F. and {Shulevski}, A. and {Toribio}, M.~C. and {ter Veen}, S. and {Zucca}, P. and {Bondonneau}, L. and {Grie{\ss}meier}, J. -M. and {Karastergiou}, A. and {Kramer}, M. and {Sobey}, C.},
        title = "{The LOFAR Tied-Array All-Sky Survey (LOTAAS): Survey overview and initial pulsar discoveries}",
      journal = {\aap},
         year = 2019,
        month = jun,
       volume = {626},
          eid = {A104},
        pages = {A104},
          doi = {10.1051/0004-6361/201935609},
archivePrefix = {arXiv},
       eprint = {1905.04977},
 primaryClass = {astro-ph.HE},
       adsurl = {https://ui.adsabs.harvard.edu/abs/2019A&A...626A.104S}
}

@ARTICLE{McEwen2020,
       author = {{McEwen}, A.~E. and {Spiewak}, R. and {Swiggum}, J.~K. and {Kaplan}, D.~L. and {Fiore}, W. and {Agazie}, G.~Y. and {Blumer}, H. and {Chawla}, P. and {DeCesar}, M. and {Kaspi}, V.~M. and {Kondratiev}, V.~I. and {LaRose}, M. and {Levin}, L. and {Lynch}, R.~S. and {McLaughlin}, M. and {Mingyar}, M. and {Noori}, H. Al and {Ransom}, S.~M. and {Roberts}, M.~S.~E. and {Schmiedekamp}, A. and {Schmiedekamp}, C. and {Siemens}, X. and {Stairs}, I. and {Stovall}, K. and {Surnis}, M. and {van Leeuwen}, J.},
        title = "{The Green Bank North Celestial Cap Pulsar Survey. V. Pulsar Census and Survey Sensitivity}",
      journal = {\apj},
         year = 2020,
        month = apr,
       volume = {892},
       number = {2},
          eid = {76},
        pages = {76},
          doi = {10.3847/1538-4357/ab75e2},
archivePrefix = {arXiv},
       eprint = {1909.11109},
 primaryClass = {astro-ph.HE},
       adsurl = {https://ui.adsabs.harvard.edu/abs/2020ApJ...892...76M}
}

@ARTICLE{Krishnan2020,
       author = {{Venkatraman Krishnan}, V. and {Flynn}, C. and {Farah}, W. and {Jameson}, A. and {Bailes}, M. and {Os{\l}owski}, S. and {Bateman}, T. and {Gupta}, V. and {van Straten}, W. and {Keane}, E.~F. and {Barr}, E.~D. and {Bhandari}, S. and {Caleb}, M. and {Campbell-Wilson}, D. and {Day}, C.~K. and {Deller}, A. and {Green}, A.~J. and {Hunstead}, R. and {Jankowski}, F. and {Lower}, M.~E. and {Parthasarathy}, A. and {Plant}, K. and {Price}, D.~C. and {Rosado}, P.~A. and {Temby}, D.},
        title = "{The UTMOST survey for magnetars, intermittent pulsars, RRATs, and FRBs - I. System description and overview}",
      journal = {\mnras},
         year = 2020,
        month = mar,
       volume = {492},
       number = {4},
        pages = {4752-4767},
          doi = {10.1093/mnras/staa111},
archivePrefix = {arXiv},
       eprint = {1905.02415},
 primaryClass = {astro-ph.IM},
       adsurl = {https://ui.adsabs.harvard.edu/abs/2020MNRAS.492.4752V}
}

@ARTICLE{Han2021,
       author = {{Han}, J.~L. and {Wang}, Chen and {Wang}, P.~F. and {Wang}, Tao and {Zhou}, D.~J. and {Sun}, Jing-Hai and {Yan}, Yi and {Su}, Wei-Qi and {Jing}, Wei-Cong and {Chen}, Xue and {Gao}, X.~Y. and {Hou}, Li-Gang and {Xu}, Jun and {Lee}, K.~J. and {Wang}, Na and {Jiang}, Peng and {Xu}, Ren-Xin and {Yan}, Jun and {Gan}, Heng-Qian and {Guan}, Xin and {Huang}, Wen-Jun and {Jiang}, Jin-Chen and {Li}, Hui and {Men}, Yun-Peng and {Sun}, Chun and {Wang}, Bo-Jun and {Wang}, H.~G. and {Wang}, Shuang-Qiang and {Xie}, Jin-Tao and {Xu}, Heng and {Yao}, Rui and {You}, Xiao-Peng and {Yu}, D.~J. and {Yuan}, Jian-Ping and {Yuen}, Rai and {Zhang}, Chun-Feng and {Zhu}, Yan},
        title = "{The FAST Galactic Plane Pulsar Snapshot survey: I. Project design and pulsar discoveries}",
      journal = {Research in Astronomy and Astrophysics},
         year = 2021,
        month = jun,
       volume = {21},
       number = {5},
          eid = {107},
        pages = {107},
          doi = {10.1088/1674-4527/21/5/107},
archivePrefix = {arXiv},
       eprint = {2105.08460},
 primaryClass = {astro-ph.HE},
       adsurl = {https://ui.adsabs.harvard.edu/abs/2021RAA....21..107H}
}

@ARTICLE{Amiri2021,
       author = {{CHIME/Pulsar Collaboration} and {Amiri}, M. and {Bandura}, K.~M. and {Boyle}, P.~J. and {Brar}, C. and {Cliche}, J. -F. and {Crowter}, K. and {Cubranic}, D. and {Demorest}, P.~B. and {Denman}, N.~T. and {Dobbs}, M. and {Dong}, F.~Q. and {Fandino}, M. and {Fonseca}, E. and {Good}, D.~C. and {Halpern}, M. and {Hill}, A.~S. and {H{\"o}fer}, C. and {Kaspi}, V.~M. and {Landecker}, T.~L. and {Leung}, C. and {Lin}, H. -H. and {Luo}, J. and {Masui}, K.~W. and {McKee}, J.~W. and {Mena-Parra}, J. and {Meyers}, B.~W. and {Michilli}, D. and {Naidu}, A. and {Newburgh}, L. and {Ng}, C. and {Patel}, C. and {Pinsonneault-Marotte}, T. and {Ransom}, S.~M. and {Renard}, A. and {Scholz}, P. and {Shaw}, J.~R. and {Sikora}, A.~E. and {Stairs}, I.~H. and {Tan}, C.~M. and {Tendulkar}, S.~P. and {Tretyakov}, I. and {Vanderlinde}, K. and {Wang}, H. and {Wang}, X.},
        title = "{The CHIME Pulsar Project: System Overview}",
      journal = {\apjs},
         year = 2021,
        month = jul,
       volume = {255},
       number = {1},
          eid = {5},
        pages = {5},
          doi = {10.3847/1538-4365/abfdcb},
archivePrefix = {arXiv},
       eprint = {2008.05681},
 primaryClass = {astro-ph.IM},
       adsurl = {https://ui.adsabs.harvard.edu/abs/2021ApJS..255....5C}
}

@ARTICLE{han06,
       author = {{Han}, J.~L. and {Manchester}, R.~N. and {Lyne}, A.~G. and {Qiao}, G.~J. and {van Straten}, W.},
        title = "{Pulsar Rotation Measures and the Large-Scale Structure of the Galactic Magnetic Field}",
      journal = {\apj},
         year = 2006,
        month = may,
       volume = {642},
       number = {2},
        pages = {868-881},
          doi = {10.1086/501444},
archivePrefix = {arXiv},
       eprint = {astro-ph/0601357},
 primaryClass = {astro-ph},
       adsurl = {https://ui.adsabs.harvard.edu/abs/2006ApJ...642..868H}
}

@ARTICLE{han18,
       author = {{Han}, J.~L. and {Manchester}, R.~N. and {van Straten}, W. and {Demorest}, P.},
        title = "{Pulsar Rotation Measures and Large-scale Magnetic Field Reversals in the Galactic Disk}",
      journal = {\apjs},
         year = 2018,
        month = jan,
       volume = {234},
       number = {1},
          eid = {11},
        pages = {11},
          doi = {10.3847/1538-4365/aa9c45},
archivePrefix = {arXiv},
       eprint = {1712.01997},
 primaryClass = {astro-ph.GA},
       adsurl = {https://ui.adsabs.harvard.edu/abs/2018ApJS..234...11H}
}

@ARTICLE{xu22,
       author = {{Xu}, Jun and {Han}, JinLin and {Wang}, PengFei and {Yan}, Yi},
        title = "{Peering into the Milky Way by FAST: III. Magnetic fields in the Galactic halo and farther spiral arms revealed by the Faraday effect of faint pulsars}",
      journal = {Science China Physics, Mechanics, and Astronomy},
         year = 2022,
        month = dec,
       volume = {65},
       number = {12},
          eid = {129704},
        pages = {129704},
          doi = {10.1007/s11433-022-2033-2},
archivePrefix = {arXiv},
       eprint = {2211.11302},
 primaryClass = {astro-ph.GA},
       adsurl = {https://ui.adsabs.harvard.edu/abs/2022SCPMA..6529704X}
}

@ARTICLE{Tyulbashev2022,
       author = {{Tyul'bashev}, Sergei A. and {Kitaeva}, Marina A. and {Tyul'basheva}, Gayane E.},
        title = "{Pushchino multibeam pulsar search - I. Targeted search of weak pulsars}",
      journal = {\mnras},
         year = 2022,
        month = nov,
       volume = {517},
       number = {1},
        pages = {1112-1125},
          doi = {10.1093/mnras/stac2404},
archivePrefix = {arXiv},
       eprint = {2203.15540},
 primaryClass = {astro-ph.HE},
       adsurl = {https://ui.adsabs.harvard.edu/abs/2022MNRAS.517.1112T}
}

@ARTICLE{Bhat2023,
       author = {{Bhat}, N.~D.~R. and {Swainston}, N.~A. and {McSweeney}, S.~J. and {Xue}, M. and {Meyers}, B.~W. and {Kudale}, S. and {Dai}, S. and {Tremblay}, S.~E. and {van Straten}, W. and {Shannon}, R.~M. and {Smith}, K.~R. and {Sokolowski}, M. and {Ord}, S.~M. and {Sleap}, G. and {Williams}, A. and {Hancock}, P.~J. and {Lange}, R. and {Tocknell}, J. and {Johnston-Hollitt}, M. and {Kaplan}, D.~L. and {Tingay}, S.~J. and {Walker}, M.},
        title = "{The Southern-sky MWA Rapid Two-metre (SMART) pulsar survey{\textemdash}II. Survey status, pulsar census, and first pulsar discoveries}",
      journal = {\pasa},
         year = 2023,
        month = may,
       volume = {40},
          eid = {e020},
        pages = {e020},
          doi = {10.1017/pasa.2023.18},
archivePrefix = {arXiv},
       eprint = {2302.11920},
 primaryClass = {astro-ph.HE},
       adsurl = {https://ui.adsabs.harvard.edu/abs/2023PASA...40...20B}
}

@ARTICLE{Tyulbashev2024,
       author = {{Tyul'bashev}, S.~A. and {Tyul'basheva}, G.~E.},
        title = "{Search for Pulsars with Periods of More Than Two Seconds at Declinations from +21{\textdegree} to +42{\textdegree}}",
      journal = {Astronomy Reports},
         year = 2024,
        month = dec,
       volume = {68},
       number = {12},
        pages = {1199-1208},
          doi = {10.1134/S1063772924701038},
       adsurl = {https://ui.adsabs.harvard.edu/abs/2024ARep...68.1199T}
}

@ARTICLE{JiangP2020RAA....20...64J,
       author = {{Jiang}, Peng and {Tang}, Ning-Yu and {Hou}, Li-Gang and {Liu}, Meng-Ting and {Kr{\v{c}}o}, Marko and {Qian}, Lei and {Sun}, Jing-Hai and {Ching}, Tao-Chung and {Liu}, Bin and {Duan}, Yan and {Yue}, You-Ling and {Gan}, Heng-Qian and {Yao}, Rui and {Li}, Hui and {Pan}, Gao-Feng and {Yu}, Dong-Jun and {Liu}, Hong-Fei and {Li}, Di and {Peng}, Bo and {Yan}, Jun and {FAST Collaboration}},
        title = "{The fundamental performance of FAST with 19-beam receiver at L band}",
      journal = {Research in Astronomy and Astrophysics},
         year = 2020,
        month = may,
       volume = {20},
       number = {5},
          eid = {064},
        pages = {064},
          doi = {10.1088/1674-4527/20/5/64},
archivePrefix = {arXiv},
       eprint = {2002.01786},
 primaryClass = {astro-ph.IM},
       adsurl = {https://ui.adsabs.harvard.edu/abs/2020RAA....20...64J}
}

@ARTICLE{rrat06,
       author = {{McLaughlin}, M.~A. and {Lyne}, A.~G. and {Lorimer}, D.~R. and {Kramer}, M. and {Faulkner}, A.~J. and {Manchester}, R.~N. and {Cordes}, J.~M. and {Camilo}, F. and {Possenti}, A. and {Stairs}, I.~H. and {Hobbs}, G. and {D'Amico}, N. and {Burgay}, M. and {O'Brien}, J.~T.},
        title = "{Transient radio bursts from rotating neutron stars}",
      journal = {\nat},
         year = 2006,
        month = feb,
       volume = {439},
       number = {7078},
        pages = {817-820},
          doi = {10.1038/nature04440},
archivePrefix = {arXiv},
       eprint = {astro-ph/0511587},
 primaryClass = {astro-ph},
       adsurl = {https://ui.adsabs.harvard.edu/abs/2006Natur.439..817M}
}

@ARTICLE{ne2001,
       author = {{Cordes}, J.~M. and {Lazio}, T.~J.~W.},
        title = "{NE2001.I. A New Model for the Galactic Distribution of Free Electrons and its Fluctuations}",
      journal = {arXiv e-prints},
         year = 2002,
        month = jul,
          eid = {astro-ph/0207156},
        pages = {astro-ph/0207156},
archivePrefix = {arXiv},
       eprint = {astro-ph/0207156},
 primaryClass = {astro-ph},
       adsurl = {https://ui.adsabs.harvard.edu/abs/2002astro.ph..7156C}
}

@ARTICLE{fast06,
       author = {{Nan}, Rendong},
        title = "{Five hundred meter aperture spherical radio telescope (FAST)}",
      journal = {Science in China: Physics, Mechanics and Astronomy},
         year = 2006,
        month = mar,
       volume = {49},
       number = {2},
        pages = {129-148},
          doi = {10.1007/s11433-006-0129-9},
       adsurl = {https://ui.adsabs.harvard.edu/abs/2006ScChG..49..129N}
}

@ARTICLE{wang2023,
       author = {{Wang}, P.~F. and {Han}, J.~L. and {Xu}, J. and {Wang}, C. and {Yan}, Y. and {Jing}, W.~C. and {Su}, W.~Q. and {Zhou}, D.~J. and {Wang}, T.},
        title = "{FAST Pulsar Database. I. Polarization Profiles of 682 Pulsars}",
      journal = {Research in Astronomy and Astrophysics},
         year = 2023,
        month = oct,
       volume = {23},
       number = {10},
          eid = {104002},
        pages = {104002},
          doi = {10.1088/1674-4527/acea1f},
archivePrefix = {arXiv},
       eprint = {2307.10340},
 primaryClass = {astro-ph.HE},
       adsurl = {https://ui.adsabs.harvard.edu/abs/2023RAA....23j4002W}
}

@ARTICLE{Zhou2023,
       author = {{Zhou}, D.~J. and {Han}, J.~L. and {Xu}, Jun and {Wang}, Chen and {Wang}, P.~F. and {Wang}, Tao and {Jing}, Wei-Cong and {Chen}, Xue and {Yan}, Yi and {Su}, Wei-Qi. and {Gan}, Heng-Qian and {Jiang}, Peng and {Sun}, Jing-Hai and {Wang}, Hong-Guang and {Wang}, Na and {Wang}, Shuang-Qiang and {Xu}, Ren-Xin and {You}, Xiao-Peng},
        title = "{The FAST Galactic Plane Pulsar Snapshot Survey. II. Discovery of 76 Galactic Rotating Radio Transients and the Enigma of RRATs}",
      journal = {Research in Astronomy and Astrophysics},
         year = 2023,
        month = oct,
       volume = {23},
       number = {10},
          eid = {104001},
        pages = {104001},
          doi = {10.1088/1674-4527/accc76},
archivePrefix = {arXiv},
       eprint = {2303.17279},
 primaryClass = {astro-ph.HE},
       adsurl = {https://ui.adsabs.harvard.edu/abs/2023RAA....23j4001Z}
}

@ARTICLE{Yao2017,
       author = {{Yao}, J.~M. and {Manchester}, R.~N. and {Wang}, N.},
        title = "{A New Electron-density Model for Estimation of Pulsar and FRB Distances}",
      journal = {\apj},
         year = 2017,
        month = jan,
       volume = {835},
       number = {1},
          eid = {29},
        pages = {29},
          doi = {10.3847/1538-4357/835/1/29},
archivePrefix = {arXiv},
       eprint = {1610.09448},
 primaryClass = {astro-ph.GA},
       adsurl = {https://ui.adsabs.harvard.edu/abs/2017ApJ...835...29Y}
}

@ARTICLE{Andrianov2025,
       author = {{Andrianov}, S.~A. and {Potapov}, V.~A. and {Tyul'bashev}, S.~A. and {Logvinenko}, S.~V. and {Oreshko}, V.~V.},
        title = "{Pushchino multibeam pulsar search VI. Method of pulsar timing using bad timed data}",
      journal = {arXiv e-prints},
         year = 2025,
        month = apr,
          eid = {arXiv:2504.08478},
        pages = {arXiv:2504.08478},
          doi = {10.48550/arXiv.2504.08478},
archivePrefix = {arXiv},
       eprint = {2504.08478},
 primaryClass = {astro-ph.HE},
       adsurl = {https://ui.adsabs.harvard.edu/abs/2025arXiv250408478A}
}

@ARTICLE{Hobbs2006,
       author = {{Hobbs}, G.~B. and {Edwards}, R.~T. and {Manchester}, R.~N.},
        title = "{TEMPO2, a new pulsar-timing package - I. An overview}",
      journal = {\mnras},
         year = 2006,
        month = jun,
       volume = {369},
       number = {2},
        pages = {655-672},
          doi = {10.1111/j.1365-2966.2006.10302.x},
archivePrefix = {arXiv},
       eprint = {astro-ph/0603381},
 primaryClass = {astro-ph},
       adsurl = {https://ui.adsabs.harvard.edu/abs/2006MNRAS.369..655H}
}

@ARTICLE{Han2025,
       author = {{Han}, J.~L. and {Zhou}, D.~J. and {Wang}, C. and {Su}, W.~Q. and {Yan}, Yi and {Jing}, W.~C. and {Yang}, Z.~L. and {Wang}, P.~F. and {Wang}, T. and {Xu}, J. and {Cai}, N.~N. and {Sun}, J.~H. and {Yang}, Q.~L. and {Xu}, R.~X. and {Wang}, H.~G. and {You}, X.~P.},
        title = "{The FAST Galactic Plane Pulsar Snapshot Survey. VI. The Discovery of 473 New Pulsars}",
      journal = {Research in Astronomy and Astrophysics},
         year = 2025,
        month = jan,
       volume = {25},
       number = {1},
          eid = {014001},
        pages = {014001},
          doi = {10.1088/1674-4527/ada3b7},
archivePrefix = {arXiv},
       eprint = {2411.15961},
 primaryClass = {astro-ph.HE},
       adsurl = {https://ui.adsabs.harvard.edu/abs/2025RAA....25a4001H}
}

@BOOK{Lorimer2004,
       author = {{Lorimer}, D.~R. and {Kramer}, M.},
        title = "{Handbook of Pulsar Astronomy}",
         year = 2004,
       volume = {4},
       adsurl = {https://ui.adsabs.harvard.edu/abs/2004hpa..book.....L}
}

\end{document}